\documentclass[%
 preprint,
 singlecolumn,
 amsmath,amssymb,
aps,
longbibliography,
showpacs]{revtex4-1}

\usepackage{graphicx}
\usepackage{dcolumn}
\usepackage{bm}
\usepackage{subfigure}
\usepackage{float}
\usepackage{color}
\usepackage{array}
\usepackage{float}
\usepackage{multirow}
\usepackage{epstopdf}
\usepackage[utf8]{inputenc}
\usepackage[T1]{fontenc}
\usepackage{mathptmx}
\usepackage{etoolbox}
\usepackage{hyperref}
\usepackage{xcolor}
\usepackage[dvipsnames]{xcolor}
\usepackage[most]{tcolorbox}

\definecolor{MyBlue}{RGB}{0,70,140}

\makeatletter
\def\@email#1#2{%
	\endgroup
	\patchcmd{\titleblock@produce}
	{\frontmatter@RRAPformat}
	{\frontmatter@RRAPformat{\produce@RRAP{*#1\href{mailto:#2}{#2}}}\frontmatter@RRAPformat}
	{}{}
}%
\makeatother



\begin{document}


\title{A Linear Time-Variant Rheological Model for Frictional Aging, Stress Relaxation, and Creep}

\author{Vikash Pandey}

\email{vikash4exploring@gmail.com}
\affiliation{%
School of Computing and Data Sciences\\FLAME University, Pune, 412115, Maharashtra, INDIA
}%
\setlength{\fboxsep}{8pt}


\begin{abstract}
Most materials undergo aging, leading to time-dependent evolution of their mechanical properties. This aging is reflected in their mechanical response to external strain and stress, which often exhibits logarithmic stress relaxation and power-law creep. Such responses are typically described using complex phenomenological models, including fractional viscoelastic formulations. While these approaches successfully reproduce experimental trends, they typically provide limited insight into the physical origin of aging and its connection to material parameters. We introduce \textit{jerk-elasticity}, a linear time-variant rheological model in which the constitutive response incorporates the time evolution of stress-rate dynamics through time-dependent material parameters. The framework is motivated by the physics of interfacial stick–slip dynamics underlying frictional aging, together with thermodynamic considerations. An asymptotic correspondence is found between the jerk-elasticity model and the rate-and-state friction law, thereby linking rheological aging with interfacial frictional aging. The proposed model reproduces the Guiu–Pratt law of logarithmic stress relaxation and Andrade's power-law creep. It further provides a framework for interpreting different creep regimes through the evolution of material parameters, without invoking distributed relaxation spectra or nonlinear constitutive assumptions. The governing parameters admit interpretation in terms of thermodynamic quantities, including activation volume, whose evolution provides a physically interpretable measure of aging. In appropriate asymptotic limits, the model recovers behaviors analogous to classical viscous and fractional Maxwell models, while also approaching Mittag–Leffler-type relaxation and Lomnitz-type creep in a specific limit. Overall, jerk-elasticity provides a minimal linear time-variant framework that links frictional aging to macroscopic creep and stress relaxation, offering an alternative to conventional phenomenological models.

\end{abstract}

\maketitle

\begin{center}
	\begin{tcolorbox}[
		width=0.94\textwidth,
		colback=MyBlue!3,
		colframe=MyBlue,
		boxrule=1pt,
		arc=1.5mm,
title=Publication Notice,
		colbacktitle=MyBlue,
		coltitle=white,
		fonttitle=\Large\bfseries,
				]
		
		\small
		
This manuscript is the author's e-print of the peer-reviewed article
published in the \emph{Journal of Rheology}. The article was selected as a
\textcolor{MyBlue}{\textbf{Featured Article by the Editors}} and appears in
\emph{Journal of Rheology}, \textbf{70}(5), 949--964 (2026).
				
\par\smallskip

	\textcolor{MyBlue}{\textbf{DOI: }}
	\href{https://doi.org/10.1122/8.0001179}
	{\nolinkurl{10.1122/8.0001179}}.

\par\smallskip
		
	\textcolor{MyBlue}{\textbf{Version of Record: }}
	\href{https://pubs.aip.org/sor/jor/article-abstract/70/5/949/3397849/A-linear-time-variant-rheological-model-for}
	{Available via AIP Publishing}.

\par\smallskip
		
	\textcolor{MyBlue}{\textbf{This e-print may differ from the Version of Record in pagination, formatting, figure sizing, referencing style, and typographic details.}}
		
\par\smallskip
{\color{MyBlue!60}\rule{\linewidth}{0.3pt}}
\par\smallskip
		
\textcolor{MyBlue}{\textbf{AIP Scilight article:}}
	
	\href{https://pubs.aip.org/aip/sci/article/2026/28/281101/3397825/Probing-the-memory-of-materials}
{\emph{Probing the memory of materials}}.

\par\smallskip
		
	    \textcolor{MyBlue}{\textbf{DOI: }}
		\href{https://doi.org/10.1063/10.0044403}
		{\nolinkurl{10.1063/10.0044403}}.

	\end{tcolorbox}
\end{center}


\section{INTRODUCTION}\label{sec1}

Aging in materials refers to the evolution of material properties that govern their mechanical response to external stress and strain \cite{Fielding2000,Siebenbuerger2012,Varchanis2019,Agarwal2019,Poling-Skutvik2020,Hem2021,Makinen2023,Farain2023}. Aging commonly manifests as slow stress relaxation, power-law creep, and the emergence of multiple deformation stages under sustained loading. As materials age, they exhibit memory \cite{Dullaert2005}, i.e., they retain memory of past stress and strain histories. Examples include disordered systems \cite{Liu2021a}, polymer glasses \cite{Amir2012}, Mylar sheets \cite{Lahini2017}, concrete \cite{Lee2024}, frictional interfaces \cite{Dillavou2018}, biological media \cite{Chen2021}, and granular media \cite{Barik2022}. Memory-dependent behavior is commonly probed through relaxation and creep tests, from which material response functions (MRFs) \cite{Mainardi2010}, namely the relaxation modulus $G\left(t\right)$ and the creep compliance $J\left(t\right)$ are defined by:
\begin{equation}
G\left(t\right) = \frac{\sigma\left(t\right)}{\varepsilon_0}, \quad
J\left(t\right) = \frac{\varepsilon\left(t\right)}{\sigma_0},
\label{simple}
\end{equation}
where $\sigma$ is stress, $\varepsilon$ is strain, and $t$ is time. Here, $\varepsilon_0$ and $\sigma_0$ denote the applied step-strain and step-stress in relaxation and creep experiments, respectively. Despite variations in material complexity arising from constituents, heterogeneity, and disorder, the functional forms of MRFs often exhibit universal features. 

Under constant strain, the stress decays slowly in materials, following the Guiu–Pratt law for logarithmic relaxation \cite{Guiu1964,Huisman2006,Aliotta2022},
\begin{equation}
\sigma\left(t\right) = \sigma_{0} - \beta \ln\left(1 + \frac{t}{\tau_{\sigma}}\right), \quad \text{where } \beta = \frac{RT}{V^{*}},
\label{gplaw}
\end{equation}
where $\tau_{\sigma}$ is the characteristic relaxation time, $V^{*}$ is the activation volume, $R$ is the universal gas constant, and $T$ is the absolute temperature. In frictional systems, the activation volume is interpreted as the characteristic volume of material that must be rearranged or activated to overcome an energy barrier, thereby enabling sliding or plastic deformation under the combined effects of stress and thermal fluctuations \cite{Long2018}.

Under constant stress, Andrade’s power-law creep \cite{Agarwal2019,Pandey2023},
\begin{equation}
\varepsilon\left(t\right)= \varepsilon_{0}\left(1 + \frac{t}{\tau_{\varepsilon}}\right)^{\alpha},
\label{andradelaw}
\end{equation}
is observed in metals \cite{Andrade1910}, fibers \cite{Nechad2005}, and particulate suspensions \cite{Kusuma2021}. In addition, Lomnitz’s logarithmic creep law,
\begin{equation}
\varepsilon_{L}\left(t\right) = \varepsilon_{0}\left[1 + \alpha_{L}\ln\left(1 + \frac{t}{\tau_{L}}\right)\right],
\label{lomnitzlaw}
\end{equation}
is observed in rocks \cite{Lomnitz1956,Pandey2016a}. Here, $\tau_{\varepsilon}$ and $\tau_{L}$ are characteristic retardation times, while $\alpha$ and $\alpha_{L}$ are material exponents. The subscript $L$ denotes quantities associated with Lomnitz’s law. Although the asymptotic behavior of logarithmic and power-law functions differs significantly, a logarithmic function can approximate power-law behavior over a finite-time window. It is, therefore, of interest to examine the rheological origins of both behaviors.

Most studies describing material response functions (MRFs) can be broadly classified into two approaches: statistical and deterministic. In the statistical approach, Guiu–Pratt’s logarithmic relaxation law is often interpreted as emerging from a broad distribution of relaxation times \cite{Hem2021,Mulla2019}, associated with complex energy landscape barriers \cite{Fielding2000,Amir2012}. In polymeric systems, relaxation is frequently modeled using a series combination of an elastic spring and a nonlinear Eyring dashpot \cite{Aliotta2022}. While constitutive relations exist for both memoryless exponential relaxation and memory-dependent power-law behavior \cite{Pandey2016a}, a corresponding linear deterministic formulation that yields logarithmic relaxation has not been developed within a standard framework. This raises questions as to whether $\beta$, and consequently $V^{*}$ in Eq.~\eqref{gplaw}, can be regarded as intrinsic material parameters \cite{Ginzburg2024}. The interpretation of the activation volume has also been discussed critically in the literature \cite{Long2018}. Analogously, in the context of power-law creep, standard statistical frameworks based on waiting-time distributions have been used to rationalize the observed behavior \cite{Koivisto2007}.

The deterministic approach to MRFs can be subdivided into two classes: linear time-invariant (LTI) models and nonlinear models. Built on the inherently local nature of Newtonian calculus, several studies describe memory through hierarchical arrangements of classical viscoelastic models in fractal and ladder networks \cite{Huisman2006}. While such network-based models can reproduce broad relaxation spectra, they do not explicitly incorporate microscopic mechanisms underlying memory in materials, as they effectively rely on approximating power-law behavior through sums of weighted exponentials. Fractional viscoelastic models, in contrast, are nonlocal and offer increased flexibility in describing memory effects, but they also have statistical interpretations \cite{Aquino2004}. Although fractional models reduce the number of fitting parameters and provide mathematical convenience, they are not generally derived from first-principles, except for a very few in which fractional dynamics was linked with the underlying physical processes \cite{Pandey2016b,Pandey2022a}. 

Finally, fiber bundle models with stick–slip dynamics \cite{Makinen2020} and several competing nonlinear models have been proposed \cite{Lee2024,Chen2021,Mulla2019}. A mean-field framework motivated by statistical thermodynamics used nonlinear constitutive relations to approximate logarithmic relaxation at short time scales in glassy amorphous solids \cite{Trachenko2021}. Similarly, in the context of creep, Schapery’s nonlinear stress–strain constitutive equation introduces nonlinearity through four stress-dependent functions; however, the reproducibility and general applicability of the model have been questioned \cite{Gamby1987}. Another example is a nonlinear damage mechanics model that yields power-law creep as a general solution of Voigt-type formulations \cite{Main2000}. Nonlinear Eyring dashpot models, originally developed for fiber composites, were used to numerically capture primary creep and finite-time tertiary creep \cite{Nechad2005}. Although nonlinear frameworks offer flexibility, they do not preserve the superposition principle, which, in turn, limits their analytical tractability due to increased mathematical complexity. Moreover, material complexities make it difficult to link macroscopic behavior to microscopic dynamics, both analytically and computationally \cite{Liu2021a,Aquino2004,Lee2010a}. Therefore, linear formulations remain widely used in describing viscoelastic response, although they may not fully capture all the observed behavior. In summary, it is increasingly recognized that the universal MRFs arise from common underlying mechanisms that are largely independent of material complexity; however, a linear, deterministic description of these responses remains an open problem \cite{Lahini2017,Weiss2023}. In this work, we introduce jerk-elasticity, a linear time-variant (LTV) rheological framework designed to bridge this gap by incorporating time-dependent constitutive parameters that encode aging dynamics. Here, ``jerk'' is used in a constitutive sense (see Sec.~\ref{sec3} for a formal definition) and should not be confused with the kinematic quantity defined as the third derivative of displacement.

The remainder of the article is organized as follows. Section~\ref{sec2} summarizes the key differences between LTI and LTV models and motivates the use of the latter for describing aging in materials. In Sec.~\ref{sec3}, the jerk-elasticity model is developed, motivated by observations of stick–slip dynamics in frictional systems. Section~\ref{sec4} is divided into several subsections. It first establishes the thermodynamic admissibility of the proposed framework, then derives the Guiu–Pratt logarithmic relaxation law and its asymptotic connection to the Mittag–Leffler function and the fractional Maxwell model. The section subsequently examines energy
dissipation, derives universal creep behavior, and develops an interpretive framework for the primary, secondary, and tertiary stages of creep. Finally, Sec.~\ref{sec5} discusses the broader implications of the proposed framework.

\section{Review of LTI and LTV Systems}\label{sec2}

We briefly recall those aspects of LTI systems that are relevant to the present formulation. Most materials are assumed to exhibit linear behavior provided they are not subjected to large-amplitude external excitation. Under this assumption, the system response is described by a linear differential equation of the form \cite{Lathi2009},
\begin{equation}
\left[a_{0}\frac{d^{n}}{dt^{n}}+a_{1}\frac{d^{n-1}}{dt^{n-1}}+\cdots +a_{n}\right]y\left(t\right)=\left[b_{n-m}\frac{d^{m}}{dt^{m}}+\cdots +b_{n-1}\frac{d}{dt}+b_{n}\right]x\left(t\right),
\label{ltide}
\end{equation}
where $x\left(t\right)$ and $y\left(t\right)$ denote the input and output, respectively, and $m$ and $n$ are positive integers. In an LTI system, the coefficients $a_{i}$ and $b_{i}$ are constants. Time invariance implies that a temporal shift in the input produces a corresponding shift in the output without altering its form. For causal systems, $x\left(t\right)=0$, $\forall t<0$. The input–output relationship can be constructed using impulse responses as follows:
\begin{align}
\text{input } & \Rightarrow\text{ output}\nonumber\\
\delta\left(t\right) & \Rightarrow h\left(t\right)\text{ (impulse response)} \nonumber\\
\delta\left(t-n\triangle\tau\right) & \Rightarrow h\left(t-n\triangle\tau\right)\text{ (time-invariance)}\nonumber\\
\left[x\left(n\triangle\tau\right)\triangle\tau\right]\delta\left(t-n\triangle\tau\right)\Rightarrow & \left[x\left(n\triangle\tau\right)\triangle\tau\right]h\left(t-n\triangle\tau\right)\text{ \text{ (homogeneity)}}\nonumber \\
\underbrace{\lim_{\triangle\tau\rightarrow0}\sum_{n}x\left(n\triangle\tau\right)\delta\left(t-n\triangle\tau\right)\triangle\tau}_{x\left(t\right)} & \Rightarrow\underbrace{\lim_{\triangle\tau\rightarrow0}\sum_{n}x\left(n\triangle\tau\right)h\left(t-n\triangle\tau\right)\triangle\tau}_{y\left(t\right)}\text{ (principle of superposition), \label{new_eq}}
\end{align}
where $\delta$ is the impulse, $h$ is the corresponding impulse response, $\triangle\tau$ is the discretized
time duration of the input, and $n\triangle\tau$ is the time at which
the input is fed to the system. In the continuous limit, the output
is given by the convolution integral
\begin{equation}
y\left(t\right)=\int\limits _{0}^{\infty}x\left(\tau\right)h\left(t-\tau\right)\,d\tau=x\left(t\right)*h\left(t\right),
\label{output}
\end{equation}
where $``*"$ denotes the convolution operation. A key property of convolution is its commutativity, i.e., for any two continuous functions, $p\left(t\right)$ and $q\left(t\right)$, $p\left(t\right)*q\left(t\right)=q\left(t\right)*p\left(t\right)$. If $p$ and $q$ are chosen as a Stieltjes inverse pair such that  $\left(p*q\right)\left(t\right)=t$, then using the Laplace transform property, $\mathcal{L}\left[p\left(t\right)*q\left(t\right)\right]=\widetilde{P}\left(s\right)\widetilde{Q}\left(s\right)$, one obtains  $\widetilde{P}\left(s\right)\widetilde{Q}\left(s\right)=1/s^{2}$. Here, $\widetilde{P}\left(s\right)=\mathcal{L}\left[p\left(t\right)\right]$
and $\widetilde{Q}\left(s\right)=\mathcal{L}\left[q\left(t\right)\right]$ are the Laplace transforms of $p\left(t\right)$ and $q\left(t\right)$ in the $s$-domain, respectively.
 
Conventionally, constitutive stress--strain equations of the LTI rheological models are expressed as linear differential equations with constant coefficients. In such idealized cases, the system response at any instant depends on the current state of the input, leading to Markovian behavior and exponential responses. Hookean elastic solids and Newtonian dashpots represent limiting examples of such memoryless systems: The former exhibits instantaneous, fully recoverable deformation, while the latter displays purely viscous dissipation without storage of deformation history. The corresponding solutions satisfy Boltzmann’s fading memory principle and can be written as temporal convolutions with memory response functions (MRFs) \cite{Mainardi2010,Pritchard2017},
\begin{equation}
\sigma\left(t\right) = \dot{\varepsilon}\left(t\right)*G\left(t\right), \quad \varepsilon\left(t\right) =\dot{\sigma}\left(t\right)*J\left(t\right),
\label{MRFs}
\end{equation}
where the number of over-dots represents the order of differentiation with respect to time. Using the convolution identity, $\dot{p}\left(t\right)*q\left(t\right)=p\left(t\right)*\dot{q}\left(t\right)$, and interpreting Eq.~$\left(\ref{MRFs}\right)$ in light of Eq.~$\left(\ref{output}\right)$ from systems theory, the derivatives  $\dot{G}\left(t\right)$ and $\dot{J}\left(t\right)$ may be identified as the impulse responses associated with relaxation and creep experiments, respectively. Taking the Laplace transform of Eq.~$\left(\ref{MRFs}\right)$ yields
\begin{equation}
\widetilde{\sigma}\left(s\right) = s\widetilde{{\varepsilon}}\left(s\right)\widetilde{G}\left(s\right) \text{ and } \widetilde{\varepsilon}\left(s\right) =s\widetilde{{\sigma}}\left(s\right)\widetilde{J}\left(s\right).
\label{LMRFs}
\end{equation}
Since both relations in Eq.~$\left(\ref{MRFs}\right)$ describe the same LTI system, they must be mutually consistent in the respective Laplace domain as well. Substituting the expression for $\widetilde{\sigma}\left(s\right)$ from the first relation of Eq.~$\left(\ref{LMRFs}\right)$ into its second relation gives the reciprocal relation
\begin{equation}
\widetilde{G}\left(s\right)\widetilde{J}\left(s\right)=\frac{1}{s^2}\Longleftrightarrow \left(G*J\right)\left(t\right)=t,
\label{SMRFs}
\end{equation}
which shows that $G\left(t\right)$ and $J\left(t\right)$ form a Stieltjes inverse pair. Within the LTI framework, the two MRFs are not independent but are intrinsically coupled through this reciprocity relation, which does not allow their independent calibration. This implies that relaxation and creep responses encode equivalent information, representing the same underlying memory kernel in dual forms. It limits the flexibility in identifying constitutive mechanisms associated with distinct physical processes that may manifest differently under stress- and strain-controlled experiments. In summary, although LTI rheology yields a mathematically elegant reciprocal structure, it enforces a redundancy between creep and relaxation descriptions. This limitation motivates the need for rheological formulations that relax the strict LTI reciprocity and allow a more flexible representation of experimentally observed memory effects.

In contrast to the defining assumption of time invariance in LTI systems, most materials encountered in practical applications are inherently time-variant, i.e., their mechanical properties evolve with time \cite{Larson2019}. In addition, they exhibit memory-laden behavior in which the response depends on the past deformation history \cite{Makinen2023,Sudreau2023}. Such systems often display nonexponential responses, including power-law and logarithmic behaviors arising from underlying microstructural evolution \cite{Poling-Skutvik2020}, which itself is driven by variations in stress and temperature \cite{Mohan2014,Dinkgreve2018,Richard2020,Oelschlaeger2022}. The governing differential equation for linearly time-variant (LTV) systems retains the structure of Eq.~$\left(\ref{ltide}\right)$, but with time-dependent coefficients. Since time invariance is no longer applicable in LTV systems, an impulse applied at different instants produces distinct responses, i.e., $\delta\left(t-n\triangle\tau\right)\nRightarrow h\left(t-n\triangle\tau\right),\text{ instead, }\delta\left(t-n\triangle\tau\right)\Rightarrow h\left(t,\tau=n\triangle\tau\right)$. The system output in this case becomes (see the footnote on p.104
in Ref.~\cite{Lathi2009})
\begin{equation}
y\left(t\right)=\int\limits _{0}^{\infty}x\left(\tau\right)h\left(t,\tau\right)\,d\tau,
	\label{misc1}
\end{equation}
where $h\left(t,\tau\right)$ is the system response at time $t$ to a unit impulse applied at time $\tau$. In this framework, the standard single-variable convolution representation underlying the classical definition of MRFs in  Eq.~$\left(\ref{MRFs}\right)$ is no longer applicable. Instead, a more general, model-independent definition of MRFs, as given in Eq.~$\left(\ref{simple}\right)$, is adopted. The loss of time invariance precludes the classical convolution structure, thereby eliminating the symmetry properties associated with LTI systems. As a result, the reciprocity relation in Eq.~$\left(\ref{SMRFs}\right)$ is no longer satisfied, leading to
\begin{equation}
\widetilde{G}\left(s\right)\widetilde{J}\left(s\right)\neq\frac{1}{s^2}\Longleftrightarrow \left(G*J\right)\left(t\right)\neq t.
\label{NMRFs}
\end{equation}
This fundamental distinction motivates the development of LTV constitutive relations, which form the basis of the present work.

Mathematically, memory effects in rheology have been represented within several complementary frameworks. LTI models describe memory through convolution integrals with time-independent kernels. While mathematically convenient, this formulation enforces a stationary memory structure and is, therefore, limited in describing materials whose internal relaxation mechanisms evolve with time. Fractional-order models extend this description by incorporating nonlocal operators with power-law kernels, thereby capturing long-range temporal correlations and broad relaxation spectra \cite{Bagley1986}. For a causal function $f\left(t\right)$, the Caputo derivative is defined as \cite{Mainardi2010} $D_{t}^{\alpha}f\left(t\right)=\dot{f}\left(t\right)*t^{-\alpha}/\Gamma\left(1-\alpha\right)$, where $\Gamma\left(\cdot\right)$ is the Euler gamma function, and the order, $\alpha\in\left[0,1\right)$. Fractional derivatives thus provide a natural generalization of integer-order differentiation. They preserve key transform properties while enabling the representation of long-memory effects. However, in their standard form, fractional models typically retain time-invariant memory kernels and thus do not account for explicit temporal evolution of the material response itself. Beyond these descriptions lies a more general but less explored class of LTV systems, in which the governing parameters, and hence the effective memory kernel evolve with time. This framework naturally captures nonstationary memory effects arising from microstructural aging, restructuring, or degradation and provides a direct description of evolving relaxation and creep behavior beyond the constraints of time-invariant formulations. 

\section{A LTV Constitutive Framework Based on Jerk-Elasticity}\label{sec3}

This section develops the constitutive foundation of jerk-elasticity as a LTV rheological framework for aging materials. The central idea is that frictional aging in stick–slip systems induces a time-dependent evolution of dissipative mechanisms, which can be represented through a stress rate–strain coupling with evolving coefficients. We first associate a mechanistic connection between interfacial aging and stress relaxation and then construct a constitutive law consistent with these observations. Finally, the resulting formulation is related asymptotically to rate-and-state friction models.

\subsection{Mechanistic link between frictional aging and stress relaxation in stick–slip dynamics}

Friction remains one of the least understood physical mechanisms despite its ubiquitous presence in engineering applications. In solids, frictional deformation exhibits a complex hierarchy of temporal processes arising from intermittent stick–slip dynamics, interfacial aging, and microstructural rearrangements. Experimental evidence shows that, under external stress and strain, internal layers of many materials develop microscopically rough interfaces due to localized deformations \cite{Galaz2018}. In metallic glasses and noncrystalline metals, such deformations give rise to narrow shear bands that slide past each other \cite{Sun2016}. The actual contact between opposing surfaces is mediated by an ensemble of microcontacts, typically in the form of asperities or nanoscale junctions protruding from these layers \cite{Suhr2005,Liu2021}. As sliding proceeds, asperities from opposing surfaces intermittently stick and lock, leading to progressive local interfacial stress accumulation. Once a threshold is reached, static friction is overcome and a rapid slip event occurs. The release of these locked contacts results in the sudden emission of stored elastic energy in the form of acoustic bursts \cite{Koivisto2007}. Such emissions were historically referred to as \textit{copper-quakes} by Andrade in his studies of deforming copper wires \cite{Andrade1910}. Following slip, the system transitions to a lower kinetic friction regime, enabling transient motion until new contacts form and the cycle repeats. Similar stick–slip behavior is observed even at atomic scales, where it arises from local bond breaking and reformation \cite{Lee2010a}. This cyclic process, when coupled with gradual interfacial evolution, gives rise to time-dependent strengthening effects commonly referred to as frictional aging.

A common observation in stick–slip frictional interactions is that the interface expands under pressure. The contact area, $A$, gradually increases with time and typically follows \cite{Dillavou2018}
\begin{equation}
A\propto \ln t.
\label{area}
\end{equation}
This logarithmic growth is often attributed to indentation creep, which triggers plastic-like flow at the asperity junctions \cite{Petrova2020}. This geometric aging stems from the adhesive and ploughing interactions that are prevalent in nonmetals and many mineral systems \cite{Dieterich1978}. In contrast, in systems with minimal roughness and plasticity, such as lamellar materials \cite{Wu-Bavouzet2007}, the origin of aging is linked to interfacial bonding and chemical strengthening, which may also evolve logarithmically in time \cite{Li2018}. The evolving geometry at the interface plays an important role in frictional aging \cite{Petrova2020} and strengthening \cite{Carpenter2016,Farain2023}.  Attempts to connect atomic-scale internal friction in disordered systems to nonaffine atomic motion are reported in Ref.~\cite{Milkus2017}. Although these effects have been probed experimentally using atomic force microscopy \cite{Tian2017}, experimental access remains challenging due to multiscale roughness confined between bulk phases \cite{Petrova2020}. Theoretically, the nonlinear extensions of the Maxwell model are used to study stick–slip-induced microsliding at asperity contacts in granular media \cite{Buckingham2000}. Despite these advances, a constitutive description of frictional aging remains an open problem due to the coupled mechanical and chemical contributions. 

To illustrate a simple mechanistic route to stress relaxation, we consider an idealized situation in which an initial strain $\varepsilon_{0}$ is maintained throughout the experiment, while the contact area evolves as $A\left(t\right)=A_{0}+\kappa\ln\left(1+t/\tau\right)$, where $A_{0}$ is the
initial contact area and $\kappa>0$ and $\tau>0$ are material constants. The shift $t/\tau\to(1+t/\tau)$ ensures regular behavior at $t=0$. The corresponding average interfacial stress is then
\begin{equation}
\sigma\left(t\right)=\frac{F}{A\left(t\right)}=\frac{\sigma_{0}}{1+\frac{\kappa}{A_{0}}\ln\left(1+\frac{t}{\tau}\right)},
	\label{misc2}
\end{equation}
where $F$ is the normal force and $\sigma_{0}=F/A_{0}$ is the initial stress. In a strain-controlled situation, the initial stress is independent of the contact area in the sense that, under linear elasticity, $\sigma_{0}=E\varepsilon_{0}$, where $E$ is the elasticity modulus. In the regime where $\left(\kappa/A_{0}\right)\ln\left(1+t/\tau\right)\ll 1$, corresponding to early-to-intermediate times, a first-order Taylor expansion yields an approximate Guiu–Pratt-type logarithmic relaxation form
\begin{equation}
\sigma\left(t\right)\approx\sigma_{0}\left[1- \frac{\kappa}{A_{0}}\ln\left(1+\frac{t}{\tau}\right)\right].
\label{stressdecay}
\end{equation}
Thus, a logarithmic increase in contact area naturally leads to a logarithmic decay of stress. Such aging mechanisms are, therefore, associated with a gradual relaxation of stress as the system evolves progressively toward more stable configurations. In this interpretation, dissipation can be viewed as arising from internal structural evolution at the interface, rather than being solely attributable to conventional viscous dissipation. Although Eq.~$\left(\ref{stressdecay}\right)$ contains the ratio $\kappa/A_{0}$, this does not yet imply a direct physical dependence on the initial contact area. The functional dependence of $\kappa$ will be clarified in Section~\ref{sec4a}, where it is shown that $\kappa$ scales with $A_{0}$, leading to an area-independent relaxation amplitude when expressed in terms of the intrinsic aging parameter.

\subsection{Jerkity as a LTV dissipative mechanism}

As discussed, frictional aging mechanisms generate history-dependent stress evolution that cannot be fully captured within standard LTI viscoelastic descriptions. In particular, cyclic buildup, locking, and release of stress at contacting asperities introduce a distinct temporal structure in which the rate of change of stress, $\dot{\sigma}\left(t\right)$, rather than the stress itself, becomes the relevant dynamical variable. This naturally leads to a LTV description, in which the stress rate–strain coupling evolves with the underlying microstructural dynamics of the interface.

We assume that both relaxation and creep arise from time-variant, stick–slip-mediated dissipative processes. This contrasts with conventional viscous dissipation, which is more appropriate for describing liquids with continuous microscopic rearrangement. We,  therefore, postulate that a broad class of materials exhibiting stick–slip-induced dissipation can be described within a time-varying rheological framework, termed \textit{jerkity}. In this formulation, the constitutive response is defined through the stress-rate relation $\dot{\sigma}\left(t\right)=\lambda\left(t\right)\varepsilon\left(t\right)$. Here, $\lambda\left(t\right)>0$ is a time-dependent material parameter, while $1/\lambda\left(t\right)$ quantifies the evolving capacity for stress relaxation arising from internal-state changes associated with interfacial stick–slip dynamics. Such a representation bypasses the need for a fixed convolution kernel; instead, it encodes memory through the dynamics of the constitutive coefficient itself. It is emphasized that the term ``jerkity'' is used strictly in a constitutive sense to denote the time-dependent coupling between stress rate and strain and should not be interpreted as a kinematic higher-order derivative of displacement \cite{Pandey2023}. Instead, it characterizes the evolving rate at which elastic stress transmission adapts to microstructural rearrangements within the material. Jerkity formulation was first introduced in Ref.~\cite{Pandey2023}, which focused on Andrade's law and Omori's law of earthquake aftershocks. The physical
justification and the relaxation behavior from the jerk mechanism are yet to be fully understood.  The microscopic origin of $\lambda\left(t\right)$ and its temporal evolution from first-principles remains an open question. However, the choice of an appropriate functional form of $\lambda\left(t\right)$ is determined uniquely by reported experimental trends and phenomenological constraints, as discussed in Subsection \ref{sec3c}.

\subsection{Functional form of the time-dependent jerkity coefficient} \label{sec3c}

Let the general form be $1/\lambda\left(t\right)=f\left(t\right)$, where $f\left(t\right)$ encodes the cumulative effects of microstructural rearrangements associated with stick–slip dynamics and frictional aging. Experimental observations of interfacial evolution, in particular, the logarithmic growth of contact area and the corresponding logarithmic relaxation of stress, suggest that $f\left(t\right)$ varies slowly in time without introducing intrinsic oscillatory or exponential time scales. In the absence of a characteristic relaxation time scale dominating the evolution, the leading-order description of such slowly varying behavior can be obtained through a local expansion in time. A local expansion about $t=0$ yields the Maclaurin series $f\left(t\right)=\xi+\theta t+\mathcal{O}\left(t^{2}\right)$. Retaining only the leading-order terms is consistent with the experimentally observed slow, monotonic evolution and avoids introducing unnecessary higher-order curvature. In addition, the coefficients of the respective nonlinear terms from the Maclaurin series would unnecessarily introduce extra material parameters without bringing any merit to the model. Alternative functional forms such as power laws or exponentials are not considered here, as they introduce intrinsic time scales or curvature inconsistent with the approximately hyperbolic stress-rate behavior observed experimentally, i.e., $\dot{\sigma}(t) \sim 1/t$.  

Notably, the presence of the first-order term in the expression of $1/\lambda\left(t\right)$ is also mandated by thermodynamic considerations. A constant $\lambda$ yields questionable transient and dynamic response. If $\lambda$ is assumed a positive constant, then the stress response from Eq.~$\left(\ref{jerkeq}\right)$ to a constant strain input is a stress that grows forever linearly with time, which is physically inadmissible. Furthermore, in a tensile experiment, it predicts an expansion, $\varepsilon\left(t\right)>0$,
for an increasing tension, $\dot{\sigma}\left(t\right)>0$, but also predicts immediate contraction under decreasing stress, $\dot{\sigma}\left(t\right)<0$. This indicates physically invalid negative work, which is also verifiable
by its dynamic response to the standard loading test. For a cyclic input of a succession of positive and negative deformations,
$\varepsilon\left(t\right)=\sin\left(\omega t\right)$, of frequency,
$\omega$, and unit amplitude, the corresponding stress response is
$\sigma\left(t\right)=\lambda\left[1-\cos\left(\omega t\right)\right]/\omega\geq0$,
which remains non-negative for all time and is, therefore, physically inconsistent with oscillatory loading. These inconsistencies stay persistent even with a negative constant value of $\lambda$, but in that case, the respective inequalities have their signs reversed. 

Finally, adopting a minimal linear form of $1/\lambda\left(t\right)$ consistent with physical constraints, the constitutive equation for the jerkity model is
\begin{equation}
\dot{\sigma}\left(t\right)=\lambda\left(t\right)\varepsilon\left(t\right), \text{ where, } \frac{1}{\lambda\left(t\right)} = \xi+\theta t.
\label{jerkeq}
\end{equation}
The coefficient of
jerkity $1/\lambda\left(t\right)$ is a time-dependent constitutive function with two parameters $\xi>0$ and $\theta>0$. For a constant strain input, integration of Eq.~$\left(\ref{jerkeq}\right)$ yields $\sigma\left(t\right)\sim \ln \left(1+\theta t/\xi\right)$. It means that considering jerkity as a purely dissipative mechanism, it reproduces the experimentally observed logarithmic stress relaxation, Eq.~$\left(\ref{stressdecay}\right)$. Accordingly, we define $\lambda\left(t\right)$ as an effective dissipative coefficient, since it governs the rate at which stress relaxes under deformation. In this view, the jerkity term may be interpreted as an effective representation of slow dissipative processes associated with structural or geometric aging. The finding also closely relates to the Burridge-Knopoff model of multicontact friction \cite{Huisman2006}. 

\subsection{Complete jerk-elasticity constitutive model}

\begin{figure}
	\begin{centering}
		\includegraphics[width=0.9\columnwidth]{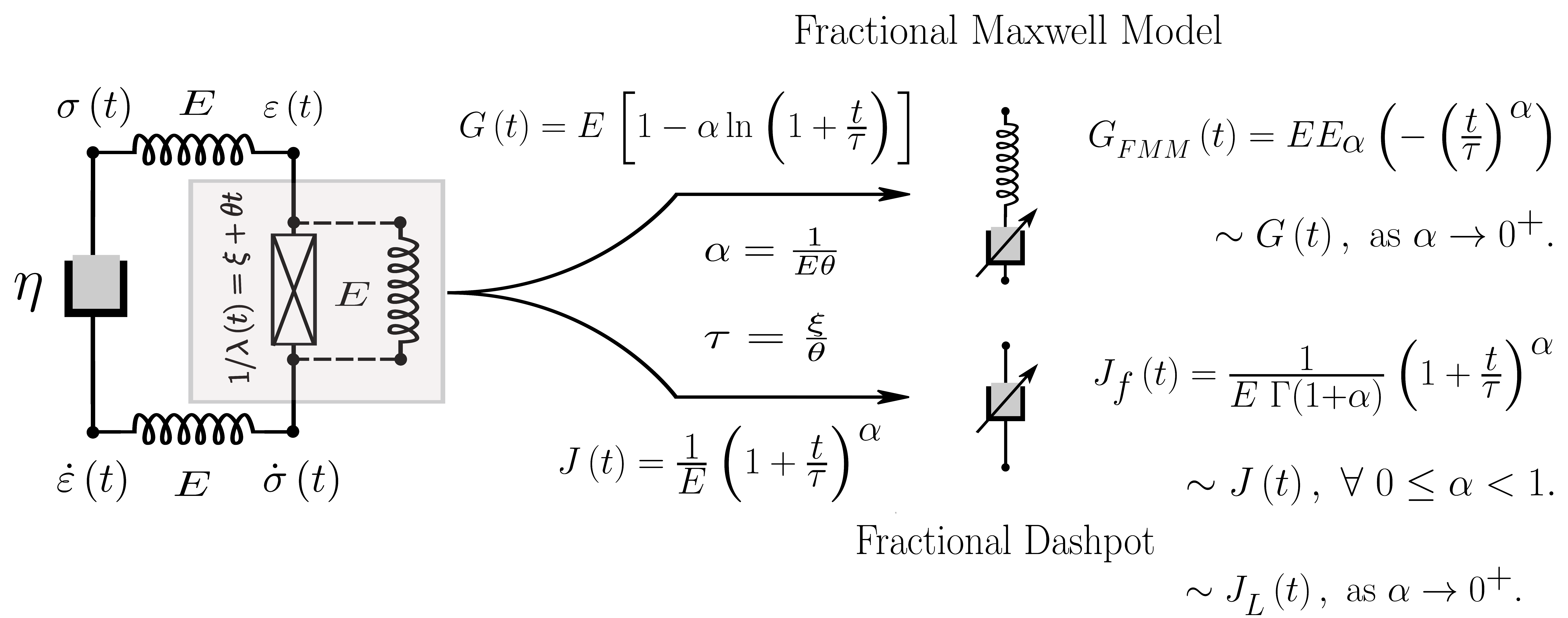}
		\par\end{centering}
	\caption{Schematic representation of the jerk-elasticity constitutive framework, in which a parallel combination of a Hookean spring (elastic modulus $E$) and a dissipative jerk element characterized by a time-dependent coefficient $\lambda\left(t\right)$ gives rise to the relation between $\dot{\sigma}\left(t\right)$ and $\varepsilon\left(t\right)$. The strain is common to both elements. The elastic branch represents energy storage, while the jerk element captures time-variant dissipative relaxation associated with stick–slip dynamics. Unlike classical linear parallel rheological networks, the two mechanisms act as competing contributions governing the evolution of the stress rate rather than independent additive stress branches. Under constant strain, the model reproduces Guiu–Pratt's logarithmic stress relaxation law. The implications extend to the fractional Maxwell model that is represented by a series combination of a spring and a fractional dashpot. As the fractional-order $\alpha\rightarrow 0^{+}$, Guiu-Pratt's law asymptotically scales as the MLF, which is inherent in the relaxation response of the fractional Maxwell model (FMM). Furthermore, the creep test on the jerk-elasticity model recovers Andrade's temporal power-law, closely approximating the response of a fractional dashpot. In the limit $\alpha\rightarrow 0^{+}$, Andrade's law approaches Lomnitz's logarithmic creep law.\\This comparison highlights that the observed logarithmic relaxation and power-law creep need not originate from LTI and fractional viscoelastic models, but may instead arise deterministically from a time-evolving constitutive law, as embodied in the jerk-elasticity model within the LTV framework.}
	\begin{centering}
		\label{JerkenFig}
		\par\end{centering}
\end{figure}

In Subsection \ref{sec3c}, we showed that both frictional aging and the jerkity mechanism give rise to logarithmic stress evolution. This correspondence suggests that the stick-phase of stick–slip dynamics, which governs interfacial aging, can be consistently represented within the jerkity framework. Since the slip-phase involves the release of the elastic energy stored during the stick-phase, it is naturally modeled by a Hookean spring with elastic modulus $E$. If all the elastic stress is dissipated through the jerkity mechanism, the stress-balance equation would be $\sigma_{s}\left(t\right)-\sigma_{d}\left(t\right)=0$, where the subscripts $s$ and $d$ denote the quantities associated with the spring and dissipative jerk, respectively. The negative sign reflects that the jerkity contribution acts as a stress-relaxing channel, i.e., a purely dissipative branch. However, in real materials, not all the stress is dissipated, instead the material retains some of the residual stress. We,  therefore, write the residual stress as $\sigma\left(t\right)=\sigma_{s}\left(t\right)-\sigma_{d}\left(t\right)$. Taking the time derivative gives $\dot{\sigma}\left(t\right)=\dot{\sigma}_{s}\left(t\right)-\dot{\sigma}_{d}\left(t\right)$. Using the constitutive relations $\sigma_{s}\left(t\right) = E\varepsilon_s\left(t\right)$ and $\dot{\sigma}_d\left(t\right) =  \lambda\left(t\right)\varepsilon_d\left(t\right)$, we obtain $\dot{\sigma}\left(t\right)=E\dot{\varepsilon}_{s}\left(t\right)-\lambda\left(t\right)\varepsilon_{d}\left(t\right)$. Assuming that both mechanisms experience the same strain field, i.e., $\varepsilon_{s}\left(t\right)=\varepsilon_{d} \left(t\right)=\varepsilon\left(t\right)$, we obtain
\begin{equation}
\dot{\sigma}\left(t\right)=E\dot{\varepsilon}\left(t\right)-\lambda\left(t\right)\varepsilon\left(t\right), \text{ where } \frac{1}{\lambda\left(t\right)} = \xi+\theta t.\label{actual_jerk}
\end{equation}
The equivalent rheological representation is schematically shown in Fig.~\ref{JerkenFig}. The system may be viewed as a generalized kinematic parallel configuration of an elastic spring and a jerk element, in which strain is shared while the stress contributions act in opposition due to competing storage and dissipation mechanisms.

In contrast to the LTI models of viscoelasticity that yield differential equations with constant coefficients, the differential equation, Eq.~$\left(\ref{actual_jerk}\right)$, has a time-varying coefficient $\lambda\left(t\right)$. Therefore, in light of Eq.~$\left(\ref{ltide}\right)$, Eq.~$\left(\ref{actual_jerk}\right)$ is an LTV system. Within the LTV framework, material memory is not encoded through a stationary convolution kernel, but through the temporal evolution of the constitutive coefficient itself. The equation represents a minimal phenomenological closure of the LTV stress-rate formulation, in which $\lambda\left(t\right)$ encodes the time-dependent dissipation arising from microstructural rearrangements associated with stick–slip dynamics. In the limit $\theta\rightarrow\infty$, the coefficient $\lambda\left(t\right)\rightarrow0$ for finite $t>0$, and Eq.~$\left(\ref{actual_jerk}\right)$ yields purely elastic response governed by Hooke’s law, $\dot{\sigma}\left(t\right) = E\dot{\varepsilon}\left(t\right)$. From here onward, the formulation expressed by Eq.~$\left(\ref{actual_jerk}\right)$ will be referred to as \textit{jerk-elasticity}. In this interpretation, $\sigma\left(t\right)$ captures the evolving stress response of the material as the internal structure reorganizes during aging, creep, and relaxation. The formulation, therefore, provides a direct bridge between microscopic interfacial dynamics and macroscopic rheological behavior.

\subsection{Asymptotic correspondence between rate-and-state friction dynamics and jerk-elasticity dynamics}

According to the rate-and-state friction (RSF) formulation of Dieterich and Ruina \cite{Dieterich1978,Pranger2022}, the shear stress associated with frictional resistance is expressed as
\begin{equation}
\sigma\left(t\right)=\sigma_r + M\ln\left(\frac{v\left(t\right)}{v_r}\right) + N\ln\left(\frac{v_r \gamma\left(t\right)}{D_c}\right),
\label{RSF}
\end{equation}
where $v\left(t\right)$ is the slip velocity, $v_r$ is a reference velocity, $\sigma_r$ is the corresponding reference stress, $D_c$ is the characteristic slip distance, $\gamma\left(t\right)$ is the internal-state variable, and $M$ and $N$ quantify the direct velocity effect and state-dependent aging contribution, respectively. The evolution of the state variable is governed by the aging law \cite{Ende2018},
\begin{equation}
\dot{\gamma}\left(t\right)=1-\frac{v\left(t\right)}{D_c}\gamma\left(t\right).
\label{aginglaw}
\end{equation}
To examine stress evolution, Eq.~$\left(\ref{RSF}\right)$ is evaluated at time, $t_{0}$. On subtracting the resulting equation from Eq.~$\left(\ref{RSF}\right)$, we have $\sigma\left(t\right)-\sigma\left(t_{0}\right)=M\ln\left[v\left(t\right)/v\left(t_{0}\right)\right]+N\ln\left[\gamma\left(t\right)/\gamma\left(t_{0}\right)\right]$, which on differentiation gives 
\begin{equation}
\dot{\sigma}\left(t\right)=M\frac{\dot{v}\left(t\right)}{v\left(t\right)} + N\frac{\dot{\gamma}\left(t\right)}{\gamma\left(t\right)}.
\label{RSF2}
\end{equation}
In a quasistatic regime corresponding to very low slip velocities, $v\gamma/D_c \ll 1$, Eq.~$\left(\ref{aginglaw}\right)$ then reduces to $\dot{\gamma}\left(t\right)\approx 1$, implying $\gamma(t)\approx \gamma_0 + t$, where $\gamma_0=\gamma\left(t_{0}\right)-t_{0}$. Under this regime, it is further reasonable to assume negligible velocity variations, i.e., $\dot{v}\left(t\right)\approx 0$. It follows from Eq.~$\left(\ref{RSF2}\right)$,
\begin{equation}
\dot{\sigma}\left(t\right)\approx \frac{N}{\gamma\left(t\right)}, \quad \text{where} \quad  \gamma(t)\approx \gamma_0 + t.
\label{RSF3}
\end{equation}
Integration yields a logarithmic stress evolution, $\sigma\left(t\right)\sim N\ln \left(1+t/\gamma_{0}\right)$, which matches the functional form obtained in studies of stick–slip-induced frictional aging and in the jerkity constitutive framework. While $N>0$ in the classical RSF formulation corresponds to frictional strengthening during interfacial aging, the same mathematical structure also admits logarithmic stress relaxation when the state-dependent contribution enters with an effective opposite sign. Such a scenario arises naturally in the jerkity framework, where dissipation governs stress decay rather than strengthening. Thus, the same mathematical structure supports both strengthening and relaxation, depending on the sign of the effective state contribution.

The correspondence becomes more transparent by noting that the RSF state variable governs interfacial aging, while the jerkity model introduces a time-dependent dissipative coefficient $\lambda\left(t\right)$. The analogous linear dependence of $1/\lambda\left(t\right)$ in Eq.~$\left(\ref{jerkeq}\right)$ and $\gamma\left(t\right)$ in Eq.~$\left(\ref{RSF3}\right)$ motivates a phenomenological correspondence,
\begin{equation} 
\gamma\left(t\right)\sim \frac{1}{\lambda\left(t\right)},
\label{correspond}
\end{equation}
which identifies both variables as internal measures of evolving contact state. Within this interpretation, $\lambda\left(t\right)$ is naturally identified as the effective dissipative coefficient, being the inverse of the jerkity coefficient introduced earlier. This correspondence admits a regime-based interpretation. For $\gamma \rightarrow \infty$, the stress rate in Eq.~$\left(\ref{RSF3}\right)$ vanishes, indicating a near-elastic, weakly dissipative limit, which corresponds to $\lambda \to 0$ in Eq.~$\left(\ref{actual_jerk}\right)$ in the jerk-elasticity framework. Physically, this corresponds to long-lived contacts and weak dissipation, approaching ideal elastic behavior. In the intermediate regime, the aging law gives $\gamma\left(t\right) \sim t$, so $\lambda\left(t\right) \sim 1/t$, leading to logarithmic stress evolution in both formulations. In the limit $\gamma \rightarrow  0$, the slip-dependent contribution in Eq.~$\left(\ref{aginglaw}\right)$ becomes negligible due to the vanishing magnitude of $\gamma$, and therefore, $\dot{\gamma} \to 1$. Within the asymptotic approximation, the expression, Eq.~$\left(\ref{RSF3}\right)$ predicts an unbounded increase in the stress rate, signaling the breakdown of the quasistatic description and the emergence of a highly sensitive, rapidly evolving, and strongly dissipative frictional state associated with freshly formed contacts. It is worth noting that, in the full rate-and-state framework, the onset of instability also depends on system stiffness and velocity-weakening parameters; the present interpretation isolates the contribution of state evolution in the quasistatic limit.  In this regime, the lack of significant state buildup weakens the stabilizing memory effect of the interface, rendering the system susceptible to unstable slip.  In the jerk-elasticity framework, this implies $\lambda \rightarrow \infty$, where Eq.~$\left(\ref{actual_jerk}\right)$, under constant stress, predicts a diverging creep rate $\dot{\varepsilon}\rightarrow \infty$, signaling instability. It will be shown in the next section that this regime corresponds to the tertiary-like creep stage leading to material failure.

\begin{table}[h!]
	\centering
		\caption{Asymptotic correspondence between RSF and jerk-elasticity.}
		\label{RSF_map}
	\resizebox{\columnwidth}{!}{%
\begin{tabular}{|c|c|c|}
	\hline
	\textbf{RSF state variable} & \textbf{Jerkity parameter} & \textbf{Physical interpretation} \\
	\hline
	$\gamma \to \infty$ & $\lambda \to 0$ & Fully aged interface; near-elastic, weakly dissipative regime \\
	\hline
	$\gamma \sim t$ & $\lambda \sim 1/t$ & Logarithmic regime; aging-dominated dynamics \\
	\hline
	$\gamma \to 0$ & $\lambda \to \infty$ & Freshly formed contacts; loss of memory; highly dissipative unstable regime \\
	\hline
\end{tabular}
	}

\end{table}

The RSF state variable $\gamma$ and the dissipative coefficient $\lambda$ in the jerk-elasticity framework, thus, provide complementary descriptions of evolving interfacial dynamics, with aging corresponding to a progressive increase in the state variable accompanied by a reduction in dissipation. The mapping in Table~\ref{RSF_map} can be visualized as a phase-like relation between interfacial aging and dissipation. As the state variable increases due to progressive interfacial aging, the dissipation coefficient decreases inversely, driving a transition from a highly dissipative, weakly aged regime toward a near-elastic, fully aged limit. This mapping identifies the jerkity coefficient as an effective inverse aging time scale, linking dissipative evolution in the LTV framework to interfacial state evolution in RSF. Importantly, just as the temporal evolution of the internal state variable $\gamma\left(t\right)$ renders the RSF formulation inherently time-variant, the evolution of $\lambda\left(t\right)$ plays an analogous role in the jerk-elasticity framework.

It is important to emphasize that this mapping should be interpreted as an asymptotic consistency rather than an exact equivalence between the two formulations. In particular, in the steady-state case $\dot{\gamma}=0$, Eq.~$\left(\ref{aginglaw}\right)$ gives $\gamma=D_{c}/v\left(t\right)$, which when substituted into Eq.~$\left(\ref{RSF}\right)$ yields  $\sigma\left(t\right)=\sigma_{r}+\left(M-N\right)\ln\left[v\left(t\right)/v_{r}\right]$. The coefficient, $M-N$, which governs velocity-strengthening or weakening behavior  does not have a direct analog within the jerk-elasticity formulation. In contrast to the RSF laws, where the stress depends explicitly on both slip rate and an internal-state variable, the jerk-elasticity framework captures aging through the temporal evolution of a constitutive coefficient $\lambda\left(t\right)$, which acts as an effective internal state. The formulation does not incorporate explicit rate dependence in the dissipative term; instead, dissipation is governed entirely by the temporal evolution of  $\lambda\left(t\right)$. In this sense, the jerk-elasticity model may be interpreted as a quasistatic or low-velocity limit of RSF, where state evolution dominates. This perspective is consistent with the emergence of logarithmic stress relaxation from slowly evolving internal structure. Nevertheless, the correspondence highlights that both descriptions encode aging through a single evolving internal quantity, differing mainly in whether it appears explicitly as a state variable in RSF or is embedded implicitly as a constitutive coefficient in the jerk-elasticity framework.

\section{Material response functions from  jerk-elasticity}\label{sec4}

This section examines the thermodynamic consistency and physical implications of the jerk-elasticity model. It first ensures admissibility and connects the formulation to the Guiu–Pratt law of logarithmic relaxation and the fractional Maxwell model, followed by an analysis of energy dissipation. The latter part addresses creep, where the framework is used to interpret the three stages of creep. The model response is also compared with classical Maxwell and Lomnitz-type descriptions.

\subsection{Thermodynamic admissibility of MRFs}\label{sec4a0}

Any set of material response functions (MRFs) derived from a physically admissible rheological model must satisfy the principles of thermodynamics and causality, thereby ensuring physically meaningful transient and dynamic behavior. In particular, for all times $t \geq 0$, the relaxation modulus $G\left(t\right)$ and creep compliance $J\left(t\right)$ must satisfy
\begin{equation}
G\left(t\right)\geq0, \text{ } \left(-1\right)^{n}\frac{d^{n}G\left(t\right)}{dt^{n}}\geq0, \quad J\left(t\right)\geq0, \text{ and }\left(-1\right)^{n}\frac{d^{n}J\left(t\right)}{dt^{n}}\leq0,
\label{Thermo_Cond}
\end{equation}
where $n$ is a positive integer \cite{Mainardi2010}. On the one
hand, the relaxation modulus is a complete monotone function, i.e.,
it is non-negative, nonincreasing, and concave upward. On the other hand,
the creep compliance is a Bernstein function, i.e., it is non-negative,
nondecreasing, and concave-downward. As the two conditions are rooted in the general
Clausius--Duhem inequality, they guarantee that the rate of entropy production
is non-negative at all times, in accordance with the second law of thermodynamics
\cite{Holm2017a}. Together, these conditions are stricter than the causality
tests motivated by Kramers--Kronig relations, as they additionally enforce passivity and thermodynamic consistency. Such thermodynamically consistent $G\left(t\right)$
and $J\left(t\right)$, when used in Boltzmann's hereditary model, correspond
to causal and fading passive memory kernels. This implies that recent loading history has a stronger influence on the current response than distant past inputs. Consequently, oscillatory behavior in the MRFs is excluded under these thermodynamically admissible conditions. 

\subsection{Guiu--Pratt's logarithmic relaxation law}\label{sec4a}

We consider a stress relaxation test in which a step strain, $\varepsilon\left(t\right)=\varepsilon_{0}$, is applied at time, $t=0$. For $t>0$, $\dot{\varepsilon}\left(t\right)=0$, and the constitutive equation, Eq.~$\left(\ref{actual_jerk}\right)$ reduces to  $\dot{\sigma}\left(t\right)=-\varepsilon_{0}/\left(\xi+\theta t\right)$,
which on integration gives  $\sigma\left(t\right)=-\left(\varepsilon_{0}/\theta\right)\ln\left(\xi+\theta t\right)+C_{\sigma}$,
where $C_{\sigma}$ is an integration constant. The initial condition is governed by the instantaneous elastic response, i.e., $\sigma\left(0\right)=\sigma_{0}=E\varepsilon_{0}$, which gives $C_{\sigma}=E\varepsilon_{0}+\left(\varepsilon_{0}/\theta\right)\ln\xi$. Substituting  $C_{\sigma}$ back into its parent expression gives 
\begin{equation}
\sigma\left(t\right)=\sigma_{0}\left[1-\alpha\ln\left(1+\frac{t}{\tau_{\sigma}}\right)\right],\text{ where } \alpha=\frac{1}{E\theta}, \text{ and }\tau_{\sigma}=\frac{\xi}{\theta}.
\label{relaxation_law}
\end{equation}
This is the Guiu-Pratt logarithmic relaxation law. Furthermore, comparison with Eq.~$\left(\ref{gplaw}\right)$ yields
\begin{equation}
\beta=\sigma_{0}\alpha=\frac{\varepsilon_{0}}{\theta}\Rightarrow\theta =\frac{\varepsilon_{0}V^{*}}{RT}.
\label{relaxation_param}
\end{equation}
Although Eq.~$\left(\ref{relaxation_law}\right)$ suggests a decomposition into an instantaneous elastic response followed by a logarithmic relaxation, this separation is only formal. The two contributions remain intrinsically coupled through the constitutive parameters. In particular, the amplitude of the logarithmic term, $-\beta=-\varepsilon_{0}/\theta$, appears independent of the elastic modulus $E$. However, the dependence on $E$ is implicitly retained through the relation $\sigma_{0}=E\varepsilon_{0}$. This coupling becomes more transparent when expressed in terms of the relaxation modulus $G\left(t\right)$, where the amplitude of the logarithmic contribution scales as $-1/\theta$. Using Eq.~$\left(\ref{relaxation_param}\right)$, this can be written as $-1/\theta=-ERT/\left(\sigma_{0}V^{*}\right)$. This expression shows that the relaxation dynamics depend not only on intrinsic material parameters but also on the initial stress level $\sigma_{0}$. Thus, the parameter $\theta$ encodes both material and state dependence, reflecting the interplay between global elastic response and local interfacial dynamics associated with stick–slip processes. Since Eq.~$\left(\ref{relaxation_param}\right)$ implies $V^{*}\sim \theta$, the activation volume is likewise both material- and state-dependent. 

The dependence of $\beta$, and consequently of $V^{*}$, on the initial deformation, $\varepsilon_{0}$, is consistent with experimental observations \cite{Ginzburg2024}. Moreover, the inverse scaling between $V^{*}$ and the initial elastic strain $\varepsilon_{0}$ is widely reported in experiments and has a clear physical interpretation. Increasing the amplitude of the initial strain, while remaining below yield, increases the initial stress and preferentially activates smaller-scale rearrangements with lower activation volumes. Consequently, the effective activation volume extracted from early-time relaxation decreases with increasing $\varepsilon_{0}$. Although $V^{*}$ may evolve slowly during relaxation, its initial value is systematically reduced at larger $\varepsilon_{0}$. This behavior supports the interpretation of $V^{*}$ as a stress-sensitive quantity, consistent with experimental findings \cite{Ginzburg2024}. 

Furthermore, the apparent $A_{0}$ dependence in Eq.~$\left(\ref{stressdecay}\right)$ is removed upon using the identification obtained from comparison with Eq.~$\left(\ref{relaxation_law}\right)$,
\begin{equation}
\kappa=\alpha A_{0}.
\label{relaxation_param_ssp}
\end{equation}
This shows that the relaxation amplitude is governed by the intrinsic aging parameter $\alpha$, while the initial contact area $A_{0}$ only sets the scaling of $\kappa$ and cancels out of the final relaxation dynamics.

The relaxation modulus $G\left(t\right)$ follows from Eq.~$\left(\ref{relaxation_law}\right)$ using its definition in Eq.~$\left(\ref{simple}\right)$, as $G\left(t\right)=E-\left(1/\theta\right)\ln\left(1+t/\tau_{\sigma}\right)$. The modulus $G\left(t\right)$ remains non-negative over the regime $t<t^{*}$, where $t^{*}=\tau_{\sigma}\left[\exp\left(1/\alpha\right)-1\right]$, beyond
which the model exhibits a breakdown which is a known characteristic of logarithmic relaxation laws. For small $\alpha$, $t^{*}$ becomes extremely large, so the
model remains valid over experimentally relevant time scales. Hence, $\forall t\leq t^{*}$,  $G\left(t\right)\geq0$. Also, $\dot{G}\left(t\right)=-\left(1/\tau_{\sigma}\theta\right)/\left(1+t/\tau_{\sigma}\right)\leq0$ and  $\ddot{G}\left(t\right)=\left(1/\tau_{\sigma}^{2}\theta\right)/\left(1+t/\tau_{\sigma}\right)^{2}\geq0$. Higher-order derivatives preserve the alternating sign structure. So, it follows from Eq.~$\left(\ref{Thermo_Cond}\right)$ that $G\left(t\right)$
is completely monotonic over experimentally relevant time scales, ensuring thermodynamic admissibility within this regime. 

The condition that $\forall t\leq t^{*}$, $G\left(t\right)\geq0$, implies that the effective bound on the logarithmic amplitude satisfies the inequality constraint, $1/\theta\leq E/\ln\left(1+t/\tau_{\sigma}\right)$.
In the limit as $t\rightarrow0^{+}$, the Maclaurin expansion leads to $\ln\left(1+t/\tau_{\sigma}\right)\approx t/\tau_{\sigma}$. 
So, $1/\theta$ stays finite as $t\rightarrow0^{+}$. In contrast,
as $t\rightarrow\infty$, $\ln\left(1+t/\tau_{\sigma}\right)\rightarrow\infty$, and the inequality becomes nonrestrictive; admissibility is then ensured by the construction requirement $\theta>0\Rightarrow 1/\theta>0$, which guarantees monotonic decay of the dissipative contribution. The same qualitative bounds apply to $\alpha$, given its relation to $\theta$ in Eq.~$\left(\ref{relaxation_law}\right)$. Physically, this indicates a relaxation process that is most pronounced at early times, where logarithmic decay is rapid, and progressively slows at longer times, consistent with an aging-driven suppression of dissipative activity in the asymptotic regime.

\subsection{Relation to fractional Maxwell model}\label{sec4a1}

The logarithmic relaxation has an inherent connection with the fractional
Maxwell model (FMM). The FMM generalizes the classical Maxwell model
by replacing the integer-order viscous element with a fractional derivative element, and it is represented as a series
combination of a spring and a fractional dashpot, as shown in Fig.~\ref{JerkenFig}. The constitutive
relation of the FMM is  $E\tau_{\sigma}^{\alpha}D_{t}^{\alpha}\varepsilon\left(t\right)=\tau_{\sigma}^{\alpha}D_{t}^{\alpha}\sigma\left(t\right)+\sigma\left(t\right)$, where $0\leq\alpha<1$, and the fractional dashpot is defined through $\sigma_{f}\left(t\right)=E\tau^{\alpha}D_{t}^{\alpha}\varepsilon_{f}\left(t\right)$. Notably, the fractional
dashpot interpolates between elastic-like and viscous-like behavior as $\alpha\rightarrow0$
and $\alpha\rightarrow1$, respectively \cite{Pritchard2017}. Imposing a relaxation test
on the FMM, we input a step-strain, and then following the applications
of the Laplace transform and correspondence principle (see Chap.~3 in Ref.~\cite{Mainardi2010}), the relaxation modulus is obtained
as 
\begin{equation}
G_{FMM}\left(t\right)=EE_{\alpha}\left(-\left(\frac{t}{\tau_{\sigma}}\right)^{\alpha}\right), \quad E_{\alpha}\left(z\right)=\sum_{k=0}^{\infty}\frac{z^{k}}{\Gamma\left(\alpha k+1\right)}\label{FMM}
\end{equation}
where $E_{\alpha}\left(z\right)$ is the Mittag-Leffler function (MLF). The MLF generalizes the exponential and plays a central role in fractional differential equations, analogous to the role of exponentials in constructing solutions to ordinary differential equations
\cite{Oliveira2011}. Since the MLF is completely monotonic, it is widely used to describe nonexponential relaxation in complex media.

At long times, $t\rightarrow\infty$, the MLF admits the asymptotic scaling, 
$E_{\alpha}\left(-t^{\alpha}\right)\sim t^{-\alpha}/\Gamma\left(1-\alpha\right)$, which yields a power-law relaxation kernel. In the same vein, $\dot{E}_{\alpha}\left(-t^{\alpha}\right)\sim-\alpha t^{-\alpha-1}/\Gamma\left(1-\alpha\right)$. In the limit of weak fractional character $\alpha\rightarrow0^{+}$, $\dot{E}_{\alpha}\left(-t^{\alpha}\right)\sim-1/t$, which on integration gives, $E_{\alpha}\left(-t^{\alpha}\right)\sim-\ln t$. Thus, as $\alpha\rightarrow0^{+}$, the power-law behavior becomes increasingly shallow, and the decay approaches a logarithmic-like regime at long time scales. Formally, this scaling crossover can be interpreted as a transition from algebraic to ultraslow relaxation, consistent with an effective logarithmic decay emerging in the limiting case.  Thus, the FMM provides a natural bridge between power-law relaxation and the logarithmic relaxation. This correspondence should be interpreted as an asymptotic consistency in relaxation scaling behavior rather than an exact equivalence between fractional and jerk-based constitutive descriptions.

\subsection{Energy dissipation in logarithmic relaxation}\label{sec4a2}

A subtle thermodynamic insight is obtained when Eqs.~$\left(\ref{relaxation_law}\right)$ and $\left(\ref{relaxation_param}\right)$ are examined together. In the expression of $\theta$ in Eq.~$\left(\ref{relaxation_param}\right)$ , the only time-dependent variable is the activation volume, $V^{*}$. Any temporal or thermal variation in $V^{*}$ is directly reflected in  $\theta$. Experimental studies have shown very weak variation in $V^{*}$ over room temperatures for materials that follow logarithmic relaxation \cite{Aliotta2022}. This can be confirmed from thermodynamic considerations as follows.

In relaxation tests, there is a rise in temperature because the material
relaxes by releasing the stored elastic energy as heat. The mechanical
power, $P$, is split into $P\left(t\right)=\dot{U}_{s}\left(t\right)+\dot{Q}_{d}\left(t\right)$, where $U_{s}$ is the stored elastic energy and $Q_{d}$
is the heat dissipation. Furthermore, $P\left(t\right)=\sigma\left(t\right)\dot{\varepsilon}\left(t\right)V$,
where $V$ is the material volume. Since in the relaxation test, $\dot{\varepsilon}\left(t\right)=0$,
we have
\begin{equation}
\dot{Q}_{d}\left(t\right)=-\dot{U}_{s}\left(t\right).
\label{heat}
\end{equation}
Under small strains and near-linear response in relaxation, the stored energy density per unit volume, $u\left(\varepsilon\right)$, is
\begin{equation}
u\left(\varepsilon\right)=\int\limits _{0}^{\varepsilon_{0}}\sigma\left(\varepsilon\right)\,d\varepsilon.
	\label{miscc3}
\end{equation}
From first principles, the incremental
mechanical work per unit volume is $\delta w=\sigma d\varepsilon$.
As all mechanical work is stored as internal energy, so $du=\delta w$,
and 
\begin{equation}
\sigma=\frac{du}{d\varepsilon}.
\label{elenergy}
\end{equation}
The fact that it is the stored elastic energy that is recovered is verifiable. As $u\left(\varepsilon\right)=E\varepsilon^{2}/2$,
then, from Eq.~$\left(\ref{elenergy}\right)$, Hooke's law, $\sigma=E\varepsilon$,
is recovered. Moreover, 
\begin{equation}
u=\frac{\sigma\varepsilon}{2}.
\label{energyd}
\end{equation}
As $U_{s}\propto u$, so, $\dot{U}_{s}\left(t\right)\propto\dot{u}\left(t\right) $, and given $\varepsilon $ is constant in relaxation tests, it follows from Eq.~$\left(\ref{energyd}\right)$, $\dot{U}_{s}\left(t\right)\propto\dot{\sigma}\left(t\right)$. With the use of the Guiu-Pratt relaxation law, Eq.~$\left(\ref{relaxation_law}\right)$, we have  $\dot{U}_{s}\left(t\right)\propto -1/\left(1+t/\tau_{\sigma}\right)$. Following, Eq.~$\left(\ref{heat}\right)$, we get $Q_{d}\left(t\right)\propto\ln \left(1+t/\tau_{\sigma}\right)$. Since temperature change, $\triangle T$, is directly proportional to the heat dissipation, we finally have
\begin{equation}
\triangle T\left(t\right) \sim \ln \left(1 + \frac{t}{\tau_{\sigma}}\right).
	\label{miscc4}
\end{equation}
This logarithmic thermal response provides a natural explanation for the weak detectability of temperature changes in materials exhibiting logarithmic stress relaxation. The near constancy of the activation volume in experiments can be interpreted as a consequence of the weak temporal variation of the dissipative channel. Logarithmic relaxation, thus, reflects a macroscopic signature of slowly evolving activation volumes associated with aging-driven interfacial rearrangements. Conversely, the model yields a falsifiable prediction: In strongly nonadiabatic conditions where heat is rapidly injected or removed, deviations from logarithmic temperature evolution are expected, leading to corresponding modifications in the effective activation volume and, therefore, in the jerkity parameter $\theta$. Such behavior is consistent with observations \cite{Aliotta2022}.

\subsection{Creep and its stages}\label{sec4b}

Complementary to stress relaxation, time-dependent deformation under constant stress provides an independent probe of aging through creep. The emergence of distinct creep-stages reflects the evolving balance between storage and dissipation mechanisms.

\subsection{Andrade's power-law as primary creep}\label{sec4b0}

The creep compliance from the jerk-elasticity model was reported in Ref.~\cite{Pandey2023}, but for completeness, we briefly reproduce it here. We apply a step-stress, $\sigma\left(t\right)=\sigma_{0}$, and determine the creep response from Eq.~$\left(\ref{actual_jerk}\right)$. We have $\dot{\sigma}_{s}\left(t\right)-\dot{\sigma}_{d}\left(t\right)=0\Rightarrow E\dot{\varepsilon}\left(t\right)-\lambda\left(t\right)\varepsilon\left(t\right)=0,$
which simplifies to $\dot{\varepsilon}\left(t\right)/\varepsilon\left(t\right)=1/\left[E\left(\xi+\theta t\right)\right]$.
On integration, we get $\ln\varepsilon\left(t\right)=\left[\ln\left(\xi+\theta t\right)\right]/\left(E\theta\right)+C_{\varepsilon}$,
where $C_{\varepsilon}$ is an integration constant. Imposing the
initial condition that all the initial stress at $t=0$ is taken by
the spring, $\sigma_{0}=\sigma_{s}=E\varepsilon_{0}$, we extract, $C_{\varepsilon}=\ln\varepsilon_{0}-\left(\ln\xi\right)/\left(E\theta\right)$.
Substituting $C_{\varepsilon}$ back into its parent expression gives
\begin{equation}
\varepsilon\left(t\right)=\varepsilon_{0}\left(1+\frac{t}{\tau_{\varepsilon}}\right)^{\alpha},\text{ }\alpha=\frac{1}{E\theta},\text{ and }\tau_{\varepsilon}=\frac{\xi}{\theta}.\label{creep}
\end{equation}
Thus, the model recovers Andrade’s power-law creep from a LTV rheological formulation, while providing a physical interpretation of the governing parameters. Andrade originally proposed that power-law creep arises from the interplay of competing deformation mechanisms \cite{Andrade1910}. Within the LTV framework, these mechanisms can be interpreted as the balance between elastic storage and jerkity-mediated dissipation. 

Furthermore, Eq.~$\left(\ref{creep}\right)$ is also mathematically equivalent to the creep compliance
of a fractional dashpot with order, $0\leq\alpha<1$. This equivalence can be verified through the
application of the Laplace transform in the constitutive relation of the fractional dashpot
\cite{Pandey2023}. The corresponding relationships are summarized in Fig.~\ref{JerkenFig}. Furthermore, combining Eqs.~$\left(\ref{relaxation_param}\right)$ and $\left(\ref{creep}\right)$, we obtain
\begin{equation}
\alpha \propto\frac{1}{V^{*}}.\label{param2}
\end{equation}
Surprisingly, the same dependence has also been reported for Lomnitz-type logarithmic creep in Ref.~\cite{Osetskii1974}.  

At all times $t\geq0$, since $J\left(t\right)\ge0$, $\dot{J}\left(t\right)=\alpha\left(1+t/\tau_{\varepsilon}\right)^{\alpha-1}/\left(E\tau_{\varepsilon}\right)\geq0$,
and $\ddot{J}\left(t\right)=\alpha\left(\alpha-1\right)\left(1+t/\tau_{\varepsilon}\right)^{\alpha-2}/\left(E\tau_{\varepsilon}^{2}\right)\leq0$, it follows that $J\left(t\right)$ satisfies the defining properties of a Bernstein-type creep function. Hence, the
jerk-elasticity mechanism expressed in Eq.~$\left(\ref{actual_jerk}\right)$
fulfills the thermodynamic admissibility conditions given in Eq.~$\left(\ref{Thermo_Cond}\right)$. The admissible range $0\le\alpha<1$ follows from the requirement $\ddot{J}\left(t\right)\le 0$, as the factor $\alpha\left(\alpha - 1\right)$ must remain nonpositive.

\begin{figure}
	\begin{centering}
		\includegraphics[width=0.8\columnwidth]{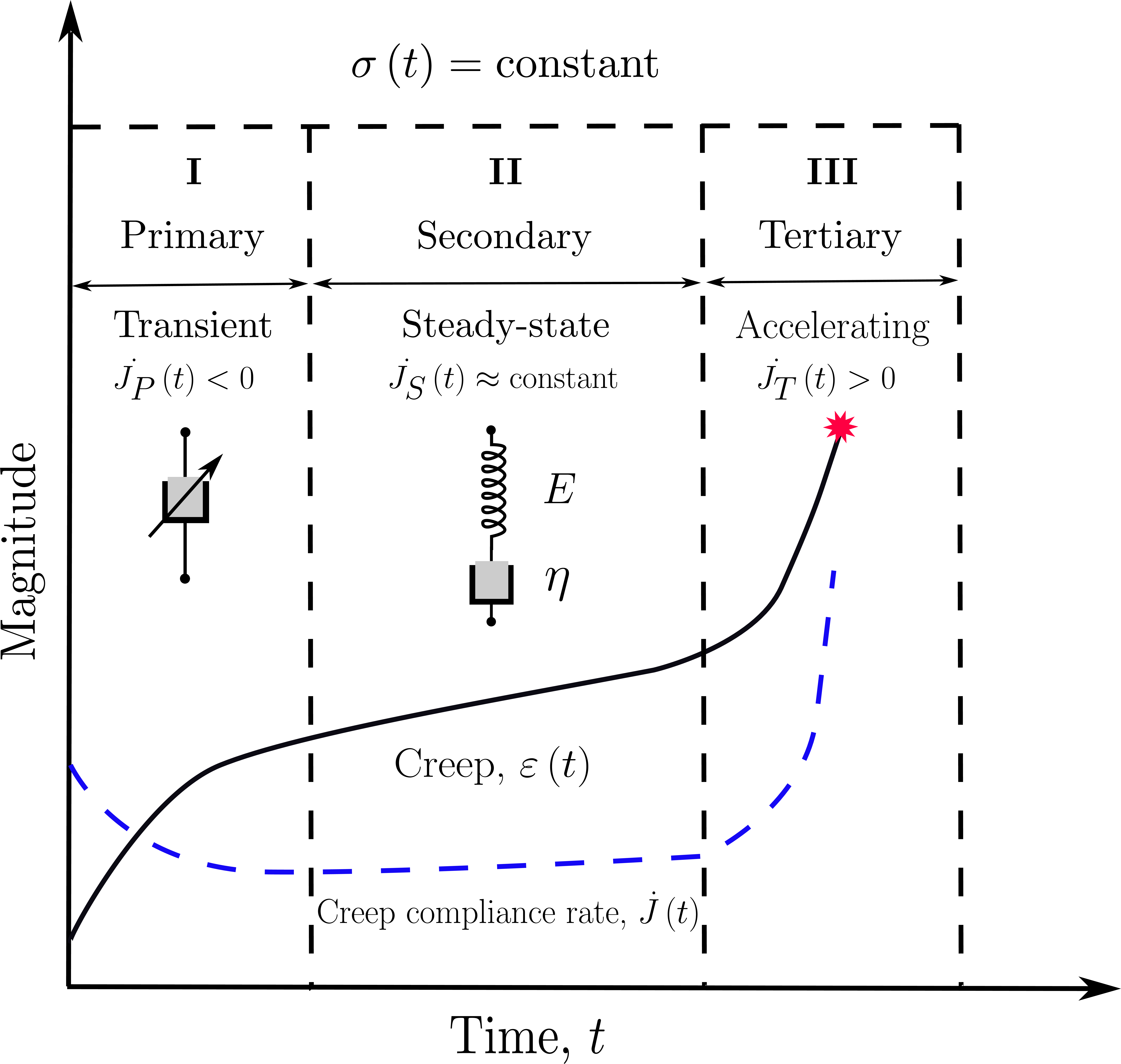}
		\par\end{centering}
	\caption{Schematic representation of the time-dependent creep response (continuous curve), $\varepsilon\left(t\right),$ and the corresponding creep-compliance rate (dashed curve), $\dot{J}\left(t\right)$,
which is proportional to the strain rate under constant applied stress. The creep response is divided into three stages: transient,
steady-state, and accelerating creep. The creep-compliance rate decreases during the transient stage, which is associated with fractional-dashpot behavior. This is followed by a steady-state regime, corresponding to a classical Maxwell-type response. In the tertiary stage, the creep-compliance rate accelerates sharply, eventually leading to material failure. This regime is not captured within the present LTV framework and lies outside its thermodynamically admissible regime.}
	\begin{centering}
		\label{CreepFig}
		\par\end{centering}
\end{figure}

Many materials exhibit a creep-compliance rate that decreases with time as $\dot{J}_{P}\left(t\right)\propto t^{\alpha-1}$, corresponding to the primary (transient) stage of deformation. This is followed by a secondary (steady-state) creep regime characterized by an approximately constant rate, $\dot{J}_{S}\left(t\right)\approx \text{constant}$. The final tertiary stage is marked by an accelerating creep rate, $\dot{J}_{T}\left(t\right)$, which increases rapidly with time and ultimately leads to material failure. This acceleration may be represented phenomenologically by a super-linear growth in time, including exponential-type behavior or finite-time divergence depending on the material system, as illustrated in Fig.~\ref{CreepFig}.

\subsection{Maxwell flow as secondary creep}\label{sec4b1}

In the limiting case $E\rightarrow 1/\theta$, it follows from Eq.~$\left(\ref{creep}\right)$, $\alpha \rightarrow 1$ and  $J\left(t\right)\propto \left(1+t/\tau_{\varepsilon}\right)$, with $\tau_{\varepsilon}=\xi/\theta\rightarrow E\xi$. 
This corresponds to a linear growth of creep compliance, characteristic of the classical Maxwell model consisting of a spring of modulus $E$ and a Newtonian dashpot of viscosity $\eta$ in series. In this sense, the secondary (steady-state) stage of creep is asymptotically equivalent to Maxwell flow. It is sometimes stated that the Maxwell model does not exhibit creep in the conventional sense, as it predicts an unbounded, linearly increasing strain without retardation. However, this interpretation reflects the absence of a mechanism that limits deformation, rather than the absence of creep itself. Within the present LTV framework, this behavior naturally emerges as the limiting case of the primary (power-law) creep regime as $\alpha \to 1$, marking a transition from retarded to steady flow. This limiting correspondence provides insight into the physical meaning of the jerkity parameters. From $\tau_{\varepsilon}=\eta/E$ in the classical Maxwell model and $\tau_{\varepsilon}=E\xi$ in the present formulation, we obtain $\xi=\eta/E^{2}$, indicating that $\xi$ carries the dimensions of time per unit modulus. Since $\lambda\left(t\right)\propto 1/\xi$, the jerkity coefficient may be interpreted as encoding the evolving efficiency of stress relaxation relative to the elastic stiffness. This supports the view that $\lambda\left(t\right)$ captures time-dependent internal restructuring that progressively facilitates flow.

At short time scales, $t \ll \tau_{\varepsilon}$, Eq.~$\left(\ref{creep}\right)$ yields $\varepsilon\left(t\right)\approx \varepsilon_{0}$, corresponding to an elastic response. At longer time scales, $t \gg \tau_{\varepsilon}$, the response approaches linear-in-time deformation, consistent with viscous flow. This crossover is consistent with the classical interpretation of viscosity as transient elasticity, originally articulated by \textit{Maxwell} in Chap.~XXI of his treatise \cite{Maxwell1871}. The transition from primary to secondary creep can, therefore, be interpreted as a shift in the dominant dissipation mechanism: from jerkity-mediated relaxation associated with evolving internal states to an effectively viscous response characterized by steady energy dissipation. 

It is worth noting that the physical mechanism responsible for the transition from elastic-like to viscous-like behavior remains an active area of research \cite{Poling-Skutvik2020,Richard2020,Sudreau2023}. Recent studies suggest that the interplay between elastic and viscous properties may not always play a governing role in phenomena such as liquid fracture \cite{Lima2026}, and alternative perspectives based on friction-induced transitions have been proposed \cite{Mari2014,Tanner2019,Lee2020,Ramaswamy2023}. In this context, the LTV framework provides a complementary interpretation in which viscous behavior emerges as an effective macroscopic response arising from the coupled evolution of elasticity and jerkity-mediated dissipation. This interpretation is consistent with the earlier correspondence between $\lambda\left(t\right)$ and the rate-and-state friction variable $\gamma\left(t\right)$, as shown in Eq.~$\left(\ref{correspond}\right)$, indicating that both creep and frictional aging are governed by a common evolving internal state.

\subsection{Unstable tertiary creep}\label{sec4b3}

The tertiary stage is typically short-lived, and is associated with accelerating deformation. This regime can be examined in the limit $\tau_{\varepsilon}/t\rightarrow\infty$, corresponding to $\theta \rightarrow 0$, where the time-variation of the rheological parameters becomes negligible compared to the evolving strain. Using Euler's definition of an exponential, i.e., in the limit
as $m\rightarrow\infty$, $\left(1+1/m\right)^{m}=\exp\left(1\right)$,
and applying it to Eq.~$\left(\ref{creep}\right)$, the tertiary
creep compliance takes the asymptotic form, $J_{T}\left(t\right)\propto\exp\left(t/\tau_{\varepsilon}\right)$. This corresponds to an accelerating, super-linear growth of strain, characteristic of tertiary creep.

In this regime, the material response departs from the slowly evolving internal-state assumption underlying the LTV framework. The resulting exponential growth reflects a loss of balance between elastic storage and dissipative relaxation, leading to unstable deformation dynamics. While this behavior may be interpreted phenomenologically as a flow regime with rapidly increasing compliance, it is not consistent with a thermodynamically admissible description. Consistently, $J_{T}\left(t\right)$ is concave upward and, therefore, does not satisfy the Bernstein-function criteria required for thermodynamic admissibility. The tertiary stage thus lies outside the consistent regime of the present formulation and is associated with material instability, typically preceding failure. This behavior contrasts with the primary and secondary regimes, which are governed by a relatively slowly evolving internal state and admit a thermodynamically consistent description within the LTV framework, thereby delineating the domain of validity of the present model.

\subsection{Interpretive framework for the three creep stages}\label{sec4b4}

Within the present LTV framework, the creep response across all three stages can be interpreted in terms of the temporal evolution of the jerk-elasticity parameter $\theta$. Specifically, $\theta$ takes relatively large values in the primary stage, approaches $\theta \rightarrow 1/E$ in the secondary stage, and tends toward $\theta \rightarrow 0$ in the tertiary stage. This progressive evolution reflects changes in the underlying internal state of the material as deformation proceeds, and may be viewed as a manifestation of aging. 

Insight into this evolution can be obtained by considering the activation volume,  $V^{*}$. Under constant applied stress, $V^{*}$ is typically observed to decrease during primary creep due to progressive strain hardening, remain approximately constant during secondary creep as a result of dynamic microstructural balance, and decrease again during tertiary creep due to localization and damage. Since both Eqs.~$\left(\ref{relaxation_param}\right)$ and $\left(\ref{param2}\right)$ suggest a direct proportionality between $V^{*}$ and $\theta$, these trends suggest a corresponding evolution of $\theta$ across the three creep stages.

The thermal response provides an additional perspective. The mechanical power density is given by $\sigma \dot{\varepsilon}\left(t\right)$, so the rate of heat generation per unit volume scales with the creep rate $\dot{\varepsilon}\left(t\right)$. In the primary stage, although the temperature may increase slightly, the rate of increase diminishes with time as the creep rate decreases, leading to rapid saturation. The secondary stage corresponds to an approximate thermal steady state, where heat generation is balanced by heat dissipation to the surroundings. In the tertiary stage, the rapid increase in $\dot{\varepsilon}\left(t\right)$ leads to significant heat generation, which may promote thermal softening and accelerate deformation, ultimately culminating in failure \cite{Hoyle2016}. Since $\dot{\varepsilon}\left(t\right)$ depends on the exponent $\alpha$, this also implies an indirect coupling between $\alpha$ and temperature through thermally activated processes. This is consistent with observations that $\alpha$ can emerge from the interplay between thermal activation and elastic stress distributions \cite{Weiss2023}.

This interpretation suggests that the evolution of $\theta$, $V^{*}$, and hence $\alpha$ provides an internal-state description linking creep, relaxation, and frictional aging within a common physical framework. Finally, while the present LTV framework provides a thermodynamically admissible description of primary and secondary creep, the tertiary stage, characterized by accelerating deformation and failure, lies outside its regime of validity. Its description requires additional mechanisms beyond those incorporated in the current formulation.

\subsection{Comparison of power-law creep with Lomnitz's logarithmic creep law}\label{sec4b5}

We compare the creep compliance predicted by Lomnitz's law, Eq.~$\left(\ref{lomnitzlaw}\right)$, with that obtained from the jerk-elasticity model. In the Lomnitz formulation, the parameters $\alpha_{_{L}}=f\left(E,\eta\left(t\right)\right)$ and $\tau_{_{L}}=g\left(\eta\left(t\right)\right)$ depend on a time-dependent viscosity $\eta(t)$, which is assumed to increase linearly with time \cite{Pandey2016a}. The motivation for this comparison arises from the observation that, despite their different underlying physical origins, both creep laws exhibit a similar asymptotic behavior: The derivative of the creep-compliance scales as $1/t$ in the limit $\alpha \rightarrow 0^{+}$.

\begin{figure}
	\begin{centering}
		\includegraphics[width=0.8\columnwidth]{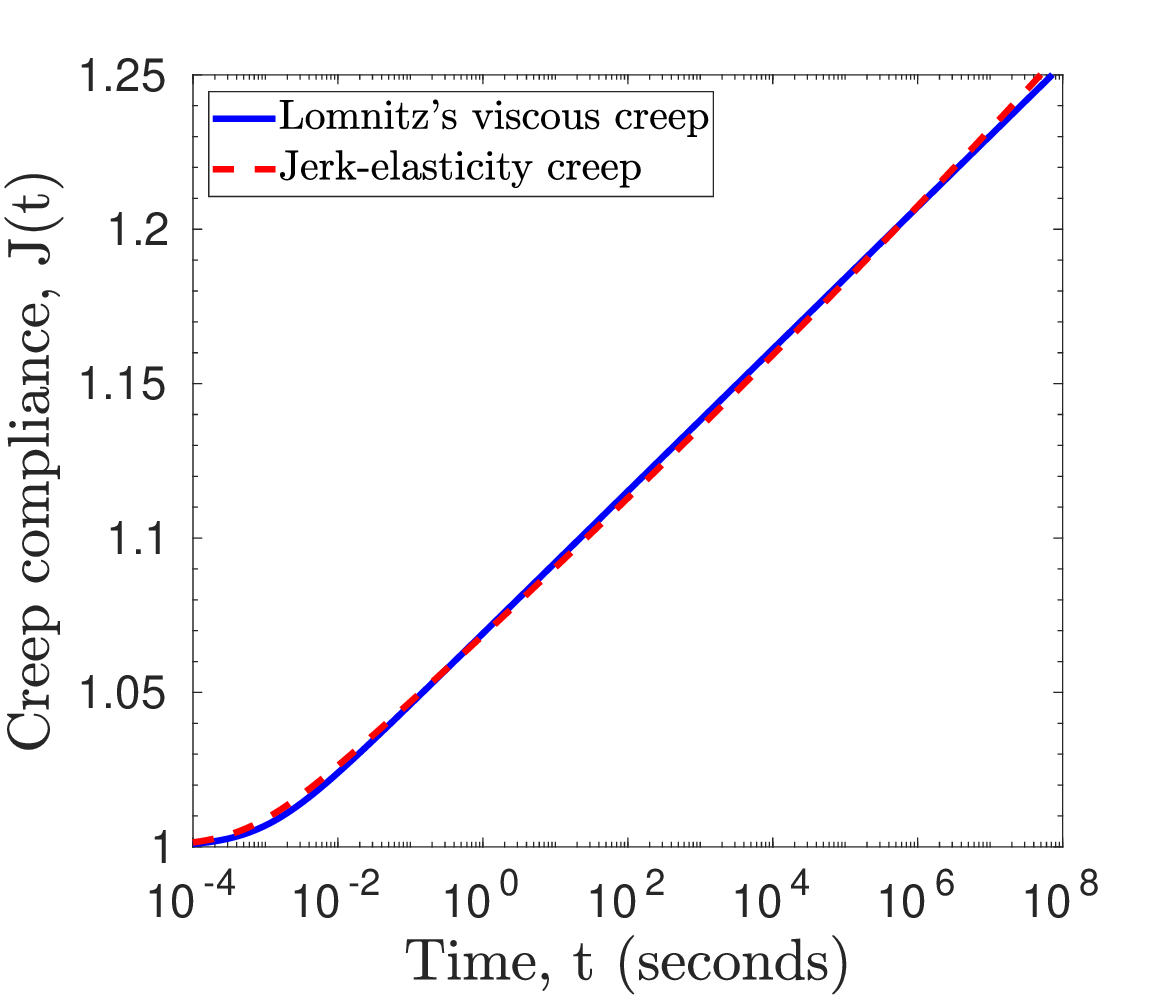}
		\par\end{centering}
	\caption{An almost perfect match between the creep compliance predicted by Lomnitz's law (continuous curve) and the jerk-elasticity model (dashed curve). The parameter values
used to obtain the plots are $\alpha\approx0.009 \ \left(0.01\right)$
and $\tau_{\varepsilon}\approx0.0006 \ \left(0.001\right)$, where the values in parentheses correspond to those obtained experimentally by Lomnitz from creep tests on igneous rocks \cite{Lomnitz1956}.}
	\begin{centering}
		\label{CreepCompFig}
		\par\end{centering}
\end{figure}

To illustrate this correspondence, we use experimentally reported values for igneous rock, $\alpha_{L}=0.01$ and $\tau_{L}=0.001$ \cite{Lomnitz1956}, to generate the creep compliance from Lomnitz's law. This response is then fitted using the creep-compliance expression from Eq.~$\left(\ref{creep}\right)$, with the elastic modulus normalized to $E=1$ in both formulations. As shown in Fig.~\ref{CreepCompFig}, an excellent agreement is obtained over a wide range of time scales, spanning from milliseconds to years. The fitted parameters, $\alpha \approx 0.009$ and $\tau_{\varepsilon} \approx 0.0006$, are in close agreement with the experimental values. The quality of the fit improves further for smaller values of $\alpha$. This correspondence indicates that the jerk-elasticity model can reproduce Lomnitz-type logarithmic creep as a limiting case, thereby providing an interpretive framework for both power-law and logarithmic creep behaviors. It also suggests that creep responses commonly attributed to time-dependent viscosity may, in certain regimes, be interpreted in terms of an evolving internal-state variable within a linear time-variant rheological description.

\section{Discussions}\label{sec5}

The present work introduces an LTV rheological framework, termed jerk-elasticity, that captures key features of aging in materials. The formulation is motivated by exploring a direct relationship between $\dot{\sigma}\left(t\right)$ and $\varepsilon\left(t\right)$ and can be interpreted more broadly as emerging from stick–slip-mediated frictional dynamics and rate-and-state descriptions. Unlike classical linear viscoelastic models with time-invariant parameters, the distinguishing feature of the present framework is the explicit temporal evolution embedded in the constitutive response.

A central outcome of the formulation is its ability to reproduce both the Guiu–Pratt logarithmic stress relaxation and Andrade-type power-law creep within a unified and interpretable framework. The rheological parameters acquire physical interpretation in terms of thermodynamic variables and activation volume. In particular, the activation volume provides a link between macroscopic deformation and microscopic energy barriers associated with structural rearrangements. Its evolution reflects progressive changes in the underlying energy landscape, consistent with aging as a gradual reconfiguration of metastable states. 

In appropriate asymptotic limits, the jerk-elasticity model recovers fractional Maxwell-type relaxation behavior. The emergence of Mittag–Leffler-type behavior further suggests a possible connection with fractional rheological models, where logarithmic and power-law memory kernels arise as limiting cases of more general internal-state dynamics.

In classical descriptions of creep, the three stages are  associated with distinct physical mechanisms. In contrast, within the present framework, these regimes emerge naturally from the temporal evolution of the internal parameters, without requiring switching between constitutive laws or introducing nonlinearities. While this does not exclude multiple mechanisms in real materials, it suggests that a significant part of the observed phenomenology may be captured through a single time-variant description. Such a unified perspective on creep stages may also provide useful insight into failure precursors in complex materials \cite{Lima2026,Lavier2021}.

Historically, creep has been associated with viscous behavior, as in Maxwell’s formulation of viscoelasticity \cite{Maxwell1871}. In conventional LTI frameworks, viscosity is introduced at the constitutive level and does not arise from conservation laws themselves. For example, the Navier–Stokes equation extends Euler’s inviscid formulation by incorporating viscosity as an additional constitutive assumption; only the latter follows directly from conservation of mass and momentum (see Chap.~1 in Ref.~\cite{Holm2019}). In contrast, the present formulation extends beyond LTI systems and shows that viscous-like behavior can emerge from a time-dependent elastic–dissipative mechanism. This perspective allows a smooth transition from elastic-like behavior in the primary creep stage to viscous-like behavior in the secondary creep stage. In this view, viscosity may be interpreted as an emergent coarse-grained manifestation of evolving internal relaxation processes, representing the limiting behavior of a more general time-dependent rheology.

An important feature of the model is that the relaxation time $\tau_{\sigma}$ and retardation time $\tau_{\varepsilon}$ share identical functional dependence, although their numerical values differ due to the evolution of the activation volume $V^{*}$. Since $V^{*}$ depends on temperature and evolves differently under creep and relaxation, the associated time scales and the exponent $\alpha$ acquire an implicit state dependence. This is consistent with experimental observations reporting distinct effective time scales in different deformation regimes \cite{Barik2022,Wang2012a}. Thus, the evolution of activation volume provides a physically interpretable measure of aging.

An intriguing result from Eq.~$\left(\ref{relaxation_param}\right)$ is that the exponent $\beta$, and consequently the Guiu–Pratt law, exhibits dependence on the initial strain $\varepsilon_{0}$. This reflects the fact that in systems with evolving internal structure, the observed macroscopic response depends not only on the constitutive form but also on the imposed loading history and initial state. While constitutive laws are typically formulated independently of boundary and initial conditions, the experimentally observed response corresponds to specific solutions of these laws, which necessarily inherit dependence on the applied constraints. In this sense, boundary conditions and initial states act as global constraints that select particular dynamical trajectories within the same constitutive framework. A similar role of boundary data arises in continuum formulations of conservation laws, where flux terms in the integral representation (e.g., in mass and charge conservation) determine the realized macroscopic evolution through the imposed external constraints.

The connection to rate-and-state friction provides additional physical grounding. The asymptotic correspondence between $\lambda\left(t\right)$ in the present model and the RSF state variable $\gamma\left(t\right)$ suggests that both descriptions encode a slowly evolving internal state governing dissipation. In this interpretation, the evolution of $\lambda\left(t\right)$ may be viewed as a macroscopic representation of contact-level processes. The proposed LTV framework may be viewed as a minimal representation of state-dependent rheology, in which the evolution of the internal structure is incorporated directly into the constitutive relation. By embedding an analogous state dependence within a rheological setting, the model bridges two domains that are often treated separately: bulk viscoelasticity and interfacial frictional dynamics. A more detailed and quantitative correspondence with specific RSF formulations, including parameter-level mapping and regime identification, remains an important direction for future work.

Despite these promising features, several limitations warrant discussion. First, although the creep and relaxation responses satisfy thermodynamic consistency conditions in the time domain, they do not satisfy the classical reciprocity relation. In particular, $\widetilde{G}\left(s\right)\widetilde{J}\left(s\right)\neq 1/s^{2}$. This follows directly from the LTV structure, for which standard Laplace-domain duality and transfer-function reciprocity are not applicable. As a result, frequency-domain analysis is not straightforward, and the formulation is most naturally interpreted in the time domain. This limitation is generic to all LTV systems rather than specific to the present model. Second, the model is linear in its constitutive structure, and thus does not explicitly account for nonlinear effects that may become significant at large deformations or high stress levels. Extending the framework to incorporate nonlinearities while preserving interpretability remains an open direction.

Third, while the jerk-elasticity model reproduces experimentally observed rheological trends and remains analytically tractable, it is fundamentally phenomenological. Although the functional form of $\lambda\left(t\right)$ is motivated by frictional aging and RSF behavior, it should be regarded as a minimal ansatz capturing the leading-order temporal evolution of the internal state. A derivation from  microstructural or statistical-mechanical principles would strengthen the theoretical foundation and clarify its range of validity. In particular, coupling the present framework with microstructural simulations could provide insight into the microscopic origin of time-variant constitutive parameters.

Fourth, the present analysis is restricted to idealized loading conditions such as step-stress and step-strain. While these are standard protocols in rheological characterization, practical applications often involve more complex histories, including cyclic or transient loading. The response of the model under such conditions remains to be investigated and may reveal additional memory-dependent features inherent to time-variant systems. Finally, although the present formulation is consistent with established empirical laws, quantitative comparison with controlled experimental datasets would be valuable for refining parameter interpretation and assessing predictive capability.

Overall, the results suggest that jerk-elasticity provides a minimal and physically transparent framework for describing aging, creep, and relaxation in materials with evolving internal structure. Rather than representing a specific constitutive law, it may be viewed as a generic representation of time-dependent rheology driven by internal-state evolution. The parsimony of the formulation, combined with its ability to interpret multiple empirical observations, supports its usefulness as a conceptual framework for studying aging in complex materials.

\begin{acknowledgments}
The author gratefully acknowledges support from the Anusandhan National Research Foundation (ANRF) under the ARG-MATRICS research funding program (ANRF/ARGM/2025/000892/MTR). The author further thanks anonymous reviewers for constructive suggestions and for prompting the investigation of links between the proposed jerk-elasticity model and rate-and-state friction (RSF) laws.
\end{acknowledgments}



\bibliographystyle{aipnum4-2}
\bibliography{Jerk_Elasticity}

\end{document}